\documentclass[12pt]{article}

\usepackage{amsthm,amsfonts,amsmath}
\usepackage[dvips]{graphicx}
\usepackage{alltt,makeidx,enumerate}

\newtheorem{thm}{Theorem}[section]
\newtheorem{lem}[thm]{Lemma}

\theoremstyle{definition}

\newtheorem{ejem}[thm]{Example}
\newtheorem{remark}[thm]{Remark}

\numberwithin{equation}{section}
\newenvironment{prueba}{\noindent\textit{Proof}:}{  \qed\\\indent}
\newcommand{\deru}[1]{#1^\prime}
\newcommand{\derd}[1]{#1^{\prime\prime}}

\newcommand{\norm}[1]{\left\Vert#1\right\Vert}

\newcommand{\set}[1]{\left\{#1\right\}}
\newcommand{\Real}{\mathbb{R}}
\newcommand{\eps}{\varepsilon}
\newcommand{\To}{\rightarrow}

\newcommand{\me}{\mathrm{e}}

\newcommand{\ts}[1]{\vec{\mathbf{#1}}}

\newcommand{\Nabla}[1]{\vec{\mathbf{\nabla}}#1}

\newcommand{\dif}{\mathrm{d}}
\newcommand{\Dif}{\mathrm{D}}

\newcommand{\Span}{\mathrm{span}}
\newcommand{\range}{\mathrm{Range}}

\newcommand{\td}[2]{\frac{\dif#1}{\dif#2}}

\newcommand{\mc}[1]{\mathcal{#1}}

\begin{document}
\title{Minimal energy configurations of finite molecular arrays}

\author{Pablo V. Negr\'on--Marrero\\
Department of Mathematics\\University of Puerto Rico\\
Humacao, PR 00791-4300\and
Melissa L\'opez--Serrano\\Department of Mathematics\\University of Puerto
Rico\\ Rio Piedras, PR 00936}
\date{}
\maketitle

\begin{abstract}
In this paper we consider the problem of characterizing the minimum energy
configurations of a finite system of particles interacting between them due to
attracting or repulsive forces given by a certain inter molecular potential. We
limit ourselves to the cases of three particles arranged in a triangular array
and that of four particles in a tetrahedral array. The minimization is
constrained to fixed area in the case of the triangular array, and to fixed
volume in the tetrahedral case. For a general class of inter molecular
potentials we give conditions for the homogeneous configuration (either an
equilateral triangle or a regular tetrahedron) of the array to be stable, that
is, a minimizer of the potential energy of the system. To determine whether or
not there exist other stable states, the system of first order necessary
conditions for a minimum is treated as a bifurcation problem with the area or
volume variable as the bifurcation parameter. Because of the symmetries present
in our problem, we can apply the techniques of equivariant bifurcation theory
to show that there exist branches of non--homogeneous solutions bifurcating from
the trivial branch of homogeneous solutions at precisely the values of the
parameter of area or volume for which the homogeneous configuration changes
stability. For the triangular array, we construct numerically the bifurcation
diagrams for both a Lennard--Jones and Buckingham potentials. The numerics
show that there exist non--homogeneous stable states, multiple stable states for
intervals of values of the area parameter, and secondary bifurcations as well.\\

  
\end{abstract}
\section{Introduction}
Consider a system of $N$ molecules, modeled as identical spherical particles,
enclosed in a region $\mc{B}$. At any given instant (or in an equilibrium
configuration), the total potential energy of the molecular array is given by:
\begin{equation}\label{basicEF}
E=\sum_{i<j}\phi(\norm{\ts{r}_i-\ts{r}_j}),
\end{equation}
where $\phi$ is the inter molecular potential energy and
$\ts{r}_1,\ldots,\ts{r}_N$ are the positions of the particles. In this paper we
consider the problem of characterizing the minimum energy configurations of this
functional in the case of three particles ($N=3$) arranged in a triangle and
that of four particles ($N=4$) in a tetrahedral array. The minimization problem
is subject to the constraint of fixed area for the triangular array and of fixed
volume in the tetrahedral case. We are particularly interested on the dependence
of the minimizing states on the parameter of area or volume in the constraint.
Both of these problems have the particularity that they can be formulated in
terms of the inter--molecular distances only, that is, without specifying the
coordinates corresponding to the positions of the particles, thus substantially
reducing the number of unknowns in each problem.

The motivation for this problem comes from the following phenomena observed
both in laboratory experiments and molecular dynamics simulations (see e.g.
\cite{BaNoSt2008}, \cite{BlKa1975}). As the density of a fluid is progressively
lowered (keeping the temperature constant), there is a certain ``critical''
density such that if the density of the fluid is lower than this critical
value, then bubbles or regions with very low density appear within the fluid.
When using discrete models of materials like \eqref{basicEF}, distinguishing
between regions of low vs high density, or whether bubbles or holes have form
within the array, is not obvious since one is dealing essentially with a set of
points. Thus to study this phenomena within this discrete model, one is
naturally led to study or characterize the stability of homogeneous energy
minimizing configurations of such an array, as the density of the the array
changes. The problems considered in this paper are the simplest problems within
such a model.

The problem of minimizing \eqref{basicEF} subject to certain type of global or
local conditions have been studied extensively (see e.g. \cite{CoKo2010},
\cite{DiBo1997}, \cite{Go1983} and the references there in). In these models
either the array is infinite, with some local repeating structure, or
finite but with $N\To\infty$. In our problem none of these conditions are
required but we expect that our results can be extrapolated to such more general
scenarios. Also we do not commit to any particular inter molecular potential
$\phi$ so that our results are applicable to any such smooth potential.

In Section \ref{sec:3p} we consider the problem of three particles. By
Heron's formula for the area of a triangle, any three numbers (representing
the inter--molecular distances) that yield a positive value for the area
formula, represent a triangle. In this case we show that the functional
\eqref{basicEF} subject to the constraint of fixed area $A$, has for any value
of $A$, a critical point representing an equilateral triangle. Moreover, in
Theorem \ref{thm:stabcond3p} we give a necessary and sufficient condition
(cf. \eqref{stab-cond}), in terms of the inter--molecular potential, for this
equilibrium point to be a (local) minimizer of the energy functional. This
condition leads to a set of values $\mc{A}$ for the area parameter $A$ for which
the equilateral triangle is a stable configuration. We give examples of how this
set looks for various inter--molecular potentials including the classical
Lennard--Jones and Buckingham potentials, and those that model hard and soft
springs including the usual Hooke's law.

Next in Section \ref{sec:nont3p} we turn to the question of whether there exists
other (not equilateral) equilibrium configurations for those values of the area
parameter $A$ for which the equilateral triangle becomes unstable, that is,
when it ceases to be a local minimizer. To answer this question, we treat the
system of equations characterizing the equilibrium points (cf.
\eqref{prob3molKT}) as a bifurcation problem with the parameter $A$ as
a bifurcation parameter, and the set of equilateral equilibrium
configurations as the trivial solution branch. We find that the necessary
condition for bifurcation from the trivial branch for this system occurs exactly
at the boundary points of the set $\mc{A}$ given by the stability condition
\eqref{stab-cond}. To check the sufficiency condition for bifurcation, one must
consider the linearization of the system \eqref{prob3molKT} about the trivial
branch at a boundary point $A_0$ of $\mc{A}$. However since the kernel of this
linearization is two dimensional we can not immediately apply the usual or
standard results from bifurcation theory. Because of the symmetries present in
this problem (cf \eqref{syms3P}), we can apply bifurcation equivariant theory to
construct a suitable reduced problem corresponding to isosceles triangular
equilibrium configurations. The linearization of the reduced problem at the
point where $A=A_0$ has now a one dimensional kernel and provided that a certain
transversality condition is satisfied (cf. \eqref{TC3p}), we can show that
there are three branches corresponding to isosceles triangles bifurcating from
the trivial branch at the point where $A=A_0$. Since the stability of these
bifurcating branches can only be determine numerically (because one must
linearized about an unknown solution), in Section \ref{sec:NE} we construct
numerically the bifurcation diagrams, with their respective stability patterns,
for instances of the Lennard--Jones and Buckingham potentials. These examples
show that the primary bifurcations off the trivial branch are of trans--critical
type, and that at least for the Lennard--Jones potential, there are secondary
bifurcations corresponding to stable scalene triangles. Moreover, there are
intervals of values of the area parameter, for which there exists
multiple stable states of the system for each value of $A$ in such an interval. 

In Section \ref{tet:sec} we consider an array of four molecules in a
tetrahedron. The general treatment in this case is similar to the three
particle case but with two main differences. First the characterization of when
six numbers (representing the lengths of the sides of the tetrahedron)
determine a tetrahedron, is given in terms of the Cayley-Menger determinant and
the triangle inequalities of one of its faces (cf. \eqref{NC_tetrah},
\eqref{cm_det}). The next complication arises from the fact that the
tetrahedron has 24 symmetries as compared to only six for the triangle! To deal
with this many possibilities, once again we make use of the basic techniques of
equivariant theory (\cite{Go2000}) to get suitable reduced problems to work
with. As the Cayley--Menger determinant is proportional to the volume of the
corresponding tetrahedron, the volume constraint in our problem is basically
that of setting this determinant to a given value $V$ for the volume. In this
case we show in Section \ref{sec:TS4p} that the functional \eqref{basicEF}
subject to the constraint of fixed volume $V$, has for any value
of $V$ a critical point representing a regular or equilateral tetrahedron.
Moreover, in Theorem \ref{triv_state_stab} we give necessary and sufficient
conditions (cf. \eqref{stab-cond_4p}), in terms of the inter--molecular
potential, for this equilibrium point to be a (local) minimizer of the energy
functional. As for the triangular case, these conditions determine a set of
values $\mc{V}$ for the volume parameter $V$ for which the equilateral
tetrahedron is a stable configuration.

In Section \ref{sec:NS4p} we consider the question of the existence of
non--equilateral equilibrium configurations. The equilibrium configurations in
this case are given as solutions to a nonlinear system of seven equations in
eight unknowns (cf. \eqref{prob4molKT}). We treat this system as a bifurcation
problem with the parameter $V$ as a bifurcation parameter, and the set of
equilateral tetrahedrons as the trivial solution branch. The necessary
condition for bifurcation from the trivial branch for this system occurs
exactly at the boundary points of the set $\mc{V}$ given by the stability
conditions \eqref{stab-cond_4p}. At a boundary point $V_0$ of $\mc{V}$ there
are two possibilities: the kernel of the linearization has dimension two or
three. Using some of the machinery of equivariance theory as in \cite{Go2000},
we can construct suitable reduced problems in each of these two cases, which
enables us to establish the existence of non-equilateral equilibrium
configurations and to get a full description of the symmetries of the
bifurcating branches (cf. Theorems \ref{ntriv4_b1}, \ref{ntriv4_b2},
\ref{ntriv4_b3}). As in the triangular case, the stability of this bifurcating
branches can only be established numerically because one must linearize about
the unknown bifurcating branch.\\

\noindent\textbf{Notation:} We let $\Real^n$ denote the $n$ dimensional space
of column vectors with elements denoted by $\ts{x}$, $\ts{y}$,$\dots$ The inner
product of $\ts{x},\ts{y}\in\Real^n$ is denoted either by
$\langle\ts{x},\ts{y}\rangle$ or $\ts{x}^t\ts{y}$, where the superscript ``$t$"
denotes transpose. We denote the set of $n\times m$ matrices by
$\Real^{n\times m}$. For $L\in\Real^{n\times m}$, we let
$\ker(L)=\{\ts{x}\in\Real^n\,:\,L\ts{x}=\ts{0}\}$ and $\range(L)=
\set{L\ts{x}\,:\,\ts{x}\in\Real^n}$. For a function
$\ts{F}\,:\,\Real^{n}\To\Real^m$, we denote
its Fr\'echet derivative by $\Dif\ts{F}$ which is given by the $m\times n$
matrix of partial derivatives of the components of $\ts{F}$. If the variables in
$\ts{F}$ are given by $(\ts{x},\ts{y})$, then $\Dif_{\ts{x}}\ts{F}$ denotes the
derivative of $\ts{F}$ with respect to the vector of variables $\ts{x}$, etc.

\section{Equivariant bifurcation from a simple eigenvalue}

In this section we provide an overview on some of the basic results on
bifurcation theory from a simple eigenvalue for mappings between finite
dimensional spaces, where the maps posses certain symmetries. The
literature on this subject is extensive but we refer to \cite{Go1983},
\cite{Am1993}, and \cite{Ki2004} for details on the material presented
in this section and further developments like for instance, the infinite
dimensional case.

Let $\ts{F}\,:\,\mc{U}\times\Real\To\Real^n$ where $\mc{U}$ is an open subset of
$\Real^{n}$, be a $C^2$ function, and
consider the problem of characterizing the solution set of:
\begin{equation}\label{bif1}
\ts{F}(\ts{x},A)=\ts{0},\quad(\ts{x},A)\in\mc{U}\times\Real.
\end{equation}
We assume that there exists a (known) smooth function $\ts{g}(\cdot)$ such that:
\[
\ts{F}(\ts{g}(A),A)=\ts{0},\quad\forall\,\,A.
\]
The set $\mc{T}=\set{(\ts{g}(A),A)\,:\,A\in\Real}$ is called the \textit{trivial
branch} of solutions of \eqref{bif1}.
We say that $(\ts{x}_0,A_0)\in\mc{T}$ is a \textit{bifurcation point} off the
trivial branch $\mc{T}$, if every neighborhood of $(\ts{x}_0,A_0)$ contains
solutions of \eqref{bif1} not in $\mc{T}$. If we let
\[
L(A)=\Dif_{\ts{x}}\ts{F}(\ts{g}(A),A),
\]
then by the Implicit Function Theorem, a necessary for $(\ts{x}_0,A_0)$ to be
a bifurcation point is that $L(A_0)$ has to be singular, a condition well known
to be not sufficient. 

In many applications of bifurcation theory and for the problems considered in
this paper, the mapping $\ts{F}$ posses symmetries due to the geometry of the
underlying physical problem. The use of these symmetries in the analysis is
useful for example to deal with problems in which $\dim\ker(L(A_0))>1$. Thus we
assume that for a proper subgroup $\mc{G}$ of $\Real^{n\times n}$,
characterizing the symmetries in the problem, the mapping
$\ts{F}$ satisfies:
\begin{equation}\label{symmF}
\ts{F}(P\ts{x},A)=P\ts{F}(\ts{x},A),\quad\forall\,P\in\mc{G}.
\end{equation}
Let $\ts{v}\in\ker(L(A_0))$ and define the \textit{isotropy subgroup} of
$\mc{G}$ at $\ts{v}$ by
\begin{equation}\label{isotsg}
\mc{H}=\set{P\in\mc{G}\,:\,P\ts{v}=\ts{v}},
\end{equation}
and the $\mc{H}$--\textit{fixed point set} by
\begin{equation}\label{fixpt}
\Real^n_{\mc{H}}=\set{\ts{x}\in\Real^n\,:\,P\ts{x}=\ts{x},\ \forall P\in\mc{H}}.
\end{equation}
Clearly $\ts{v}\in\Real^n_{\mc{H}}$. 

Let $\mathbb{P}_{\mc{H}}\,:\Real^n\To\Real^n$ be a linear map that projects
onto $\Real^n_{\mc{H}}$, that is $\range(\mathbb{P}_{\mc{H}})=\Real^n_{\mc{H}}$
and $\mathbb{P}_{\mc{H}}(\Real^n_{\mc{H}})=\Real^n_{\mc{H}}$. With
$\mc{U}_{\mc{H}}=\mathbb{P}_{\mc{H}}(\mc{U})=\mc{U}\cap\Real^n_{\mc{H}}$, we
define $\ts{F}_{\mc{H}}\,:\,\mc{U}_{\mc{H}}\times\Real \To\Real^n_{\mc{H}}$ by:
\begin{equation}\label{reducedF}
\ts{F}_{\mc{H}}(\ts{u},A)=\mathbb{P}_{\mc{H}}\ts{F}(\ts{u},A),\quad
(\ts{u},A)\in\mc{U}_{\mc{H}}\times\Real.
\end{equation}
An easy calculation now gives that
\[
\Dif_{\ts{u}}\ts{F}_{\mc{H}}(\ts{u},A)=\mathbb{P}_{\mc{H}}
\ts{F}_{\ts{x}}(\ts{u},A)\mathbb{P}_{\mc{H}}.
\]
We assume that $\ts{g}(A)\in\Real^n_{\mc{H}}$ for all $A$, so that
\[
\ts{F}_{\mc{H}}(\ts{g}(A),A)=\ts{0},\quad\forall\,\,A.
\]
It follows now that $L_{\mc{H}}(A)\,:\,\Real^n_{\mc{H}}\To\Real^n_{\mc{H}}$ is
given by:
\[
L_{\mc{H}}(A)=\Dif_{\ts{u}}\ts{F}_{\mc{H}}(\ts{g}(A),A)=
\mathbb{P}_{\mc{H}}L(A)\mathbb{P}_{\mc{H}}.
\]
Clearly $\ts{v}\in\ker(L_{\mc{H}}(A_0))$. The $\mc{H}$--\textit{reduced
problem} is now given by:
\begin{equation}\label{Hreduced}
\ts{F}_{\mc{H}}(\ts{u},A)=\ts{0},\quad
(\ts{u},A)\in\mc{U}_{\mc{H}}\times\Real.
\end{equation}
An important property relating \eqref{bif1} and \eqref{Hreduced} is that
$(\ts{x},A)\in\mc{U}_{\mc{H}}\times\Real$ is a solution of \eqref{bif1} if and
only if $(\ts{x},A)$ is a solution of \eqref{Hreduced}. The following result
provides the required sufficient conditions for $(\ts{x}_0,A_0)$ to be a
bifurcation point of the $\mc{H}$--reduced problem.

\begin{thm}[Equivariant Bifurcation Theorem \cite{He1988}]\label{Kras}
Assume that for $A=A_0$ there exists $\ts{v}\in\ker(L(A_0))$ that defines a
proper isotropy subgroup $\mc{H}$ such that:
\[
\ker(L_{\mc{H}}(A_0))=\Span\set{\ts{v}},\quad
L_{\mc{H}}^\prime(A_0)\ts{v}
\notin\range(L_{\mc{H}}(A_0)).
\]
Then there exists a branch $\mc{C}_{\mc{H}}$ of nontrivial solutions
of $\ts{F}_{\mc{H}}(\ts{u},A)=\ts{0}$ bifurcating
from the trivial branch $\mc{T}$ at the point where
$A=A_0$ and such that either:
\begin{enumerate}[i)]
\item
$\mc{C}_{\mc{H}}$ is unbounded in $\Real^{n+1}$;
\item
the closure of $\mc{C}_{\mc{H}}$ intersects the boundary $\partial \mc{U}$ of
$\mc{U}$;
\item
$\mc{C}_{\mc{H}}$ intersects $\mc{T}$ at a point $(\ts{x}_*,A_*)$ where $A_*\ne
A_0$.
\end{enumerate}
\end{thm}

The proof of this theorem is basically an application of a result from
Krasnoselski \cite{Kra1965} that uses the homotopy invariance of the topological
degree. The three alternatives in the statement of the theorem are usually
referred to as the Crandall and Rabinowitz alternatives. The local version of
this result (cf. \cite{Go1983}), that is, without the Crandall and Rabinowitz
alternatives,  can be obtained via the Lyapunov--Schmidt reduction method. A
useful consequence of this reduction is an approximate formula for the
bifurcating branch in a neighborhood of the bifurcation point. Let
$\ker(L_{\mc{H}}(A_0)^t)=\Span\set{\ts{v}^*}$, where
$\langle\ts{v}^*,\ts{v}\rangle=1$, so that $\range(L_{\mc{H}}(A_0))=
\set{\ts{y}\in\Real^n_{\mc{H}}\,:\,\langle\ts{v}^*,\ts{y}\rangle=0}$. Now if we
define
\begin{equation}\label{bif2}
A^0=\langle\ts{v}^*,(\Dif_{\ts{u}\ts{u}}\ts{F
}_{\mc{H}}^0\ts{v})\ts{v}\rangle,\quad
B^0=\langle\ts{v}^*,\deru{L}_{\mc{H}}(A_0)\ts{v}\rangle.
\end{equation}
(here the zero superscripts mean evaluated at $(\ts{x}_0,A_0)$), then the
bifurcating branch have the following asymptotic expansion: 
\begin{equation}\label{bif3}
(\ts{x},A)=\left[\ts{g}(A_0+\eps)+\eps
m\ts{v}+O(\eps^2),A_0+\eps\right],
\end{equation}
where
\begin{equation}\label{bif4}
m=-\dfrac{2B^0}{A^0},\quad A^0\ne0.
\end{equation}

\section{The three particle case}\label{sec:3p}
In this section we consider the case in which the molecular array consists
of three particles. The inter--particle energy is given by a smooth function
$\phi\,:\,(0,\infty)\To\Real$ called the \textit{potential}. If $(a,b,c)$ are
the distances between the particles in the array, the total energy of the
system is given by:
\begin{equation}\label{energy3}
E(a,b,c)=\phi(a)+\phi(b)+\phi(c),\quad a,b,c>0.
\end{equation}
Also, the square of the area of the triangular array is given by Heron's
formula:
\[
g(a,b,c)\equiv s(s-a)(s-b)(s-c),
\]
where $s=(a+b+c)/2$.

For any any given number $A>0$, we consider the following
constrained minimization problem:
\begin{equation}\label{prob3}
\left\{\begin{array}{c}\displaystyle\min_{a,b,c>0}E(a,b,c)\\\mbox{subject to
}g(a,b,c)=A^2.
\end{array}\right.
\end{equation}
Thus we are looking for minimizers of the energy functional $E$ subject to the
constraint that the area of the array is $A$. The first order necessary
conditions for a solution of this problem are given by:
\begin{equation}\label{prob3molKT}
\left\{\begin{array}{rcl} g(a,b,c)-A^2&=&0,\\\Nabla E(a,b,c)+\lambda\Nabla
g(a,b,c)&=&\ts{0},
\end{array}\right.
\end{equation}
where $\lambda\in\Real$ is the Lagrange multiplier corresponding to the
restriction $g(a,b,c)=A^2$. For any given value of $A>0$, this is a nonlinear
system of equations for the unknowns $(\lambda,a,b,c)$ in terms of $A$.
In general this system can have multiple solutions depending on the
characteristics of the potential $\phi$ and the value of $A$.
\subsection{Existence and stability of trivial states}
An easy calculation shows that:
\begin{equation}\label{gradEg1}
\Nabla
E=\left[\begin{array}{c}\deru{\phi}(a)\\\deru{\phi}(b)\\\deru{\phi}(c)
\end{array}\right],\quad
\Nabla
g=\frac{1}{4}\left[\begin{array}{c}a(b^2+c^2-a^2)\\b(a^2+c^2-b^2)\\
c(a^2+b^2-c^2)\end{array}\right].
\end{equation}
Thus the system \eqref{prob3molKT} is equivalent to:
\begin{equation}\label{prob3:1stord}
\left\{\begin{array}{rcl}
\frac{1}{8}(a^2b^2+a^2c^2+b^2c^2)-\frac{1}{16}(a^4+b^4+c^4)-A^2&=&0,\\
\deru{\phi}(a)+\frac{\lambda}{4}a(b^2+c^2-a^2)&=&0,\\
\deru{\phi}(b)+\frac{\lambda}{4}b(a^2+c^2-b^2)&=&0,\\
\deru{\phi}(c)+\frac{\lambda}{4}c(a^2+b^2-c^2)&=&0.
\end{array}\right.
\end{equation}
This system always has a solution with $a=b=c$. In fact upon setting $a=b=c$ in
\eqref{prob3:1stord}, this system reduces to:
\begin{equation}\label{tribranch}
\dfrac{3}{16}\,a^4=A^2,\quad \deru{\phi}(a)=-\dfrac{\lambda a^3}{4}.
\end{equation}
Thus we have the following result:
\begin{lem}\label{triv_state}
For any value of $A>0$, the system \eqref{prob3:1stord} has a solution of the
form\\ $(\lambda_A,a_A,a_A,a_A,A)$ where:
\begin{equation}\label{symcpt}
a_A=\dfrac{2\sqrt{A}}{\sqrt[4]{3}},\quad\lambda_A=-\dfrac{4\deru{\phi}(a_A)}{
a_A^3 } .
\end{equation}
\end{lem}

We now characterize for which values of $A$, the solution provided in Lemma
\ref{tribranch} is actually a minimizer, i.e., a solution of \eqref{prob3}. To
do this we need to examine the matrix
$\left[\nabla^2E+\lambda\nabla^2g\right](\ts{v}_A)$ where
$\ts{v}_A=(\lambda_A,a_A,a_A,a_A)$, over the subspace:
\begin{equation}\label{subspc2nd}
\mc{M}=\set{(x,y,z)\,:\,\Nabla g(\ts{v}_A)\cdot(x,y,z)=0}.
\end{equation}
A straightforward calculation gives that:
\begin{eqnarray}
\nabla^2E&=&\left[\begin{array}{ccc}\derd{\phi}(a)&0&0\\0&\derd{\phi}(b)&0\\
0&0&\derd{\phi}(c)\end{array}\right],\label{gradEq2}\\
\nabla^2g&=&\frac{1}{4}\left[\begin{array}{ccc}b^2+c^2-3a^2&2ab&2ac\\
2ab&a^2+c^2-3b^2&2bc\\2ac&2bc&a^2+b^2-3c^2\end{array}\right].\label{gradEg3}
\end{eqnarray}
It follows now that
\begin{equation}\label{Hessiansym}
\left[\nabla^2E+\lambda\nabla^2g\right](\ts{v}_A)=\left[
\begin{array}{ccc}
\derd{\phi}(a_A)-\frac{\lambda_A a_A^2}{4}&\frac{\lambda_A
a_A^2}{2}&\frac{\lambda_A a_A^2}{2}\\
\frac{\lambda_Aa_A^2}{2}&\derd{\phi}(a_A)-\frac{\lambda_A
a_A^2}{4}&\frac{\lambda_A a_A^2}{2}\\
\frac{\lambda_Aa_A^2}{2}&\frac{\lambda_Aa_A^2}{2}&
\derd{\phi}(a_A)-\frac{\lambda_A a_A^2}{4}
\end{array}\right],
\end{equation}
and that
\[
\mc{M}=\set{(x,y,z)\,:\,x+y+z=0}.
\]
Using \eqref{symcpt} one can show now that the matrix \eqref{Hessiansym} is
positive definite over $\mc{M}$ if and only if:
\begin{equation}\label{stab-cond}
\derd{\phi}(a_A)+\frac{3}{a_A}\deru{\phi}(a_A)>0.
\end{equation}
Thus we have the following result:
\begin{thm}\label{thm:stabcond3p}
Let $\phi\,:\,(0,\infty)\To\Real$ be twice continuously differentiable function.
Then the uniform array $(a,b,c)=(a_A,a_A,a_A)$ in Lemma \ref{triv_state} is a
relative minimizer for the problem \eqref{prob3} for those values of $A$ for
which \eqref{stab-cond} holds.
\end{thm}

\begin{ejem}\label{LJpot}
Consider the case of a potential that has the following form:
\begin{equation}\label{LJ-potential}
\phi(r)=\dfrac{c_1}{r^{\delta_1}}-\dfrac{c_2}{r^{\delta_2}},
\end{equation}
where $c_1,c_2$ are positive constants and $\delta_1>\delta_2>2$. (These constants determine the physical properties of the particle or molecule in question. The classical \textit{Lennard--Jones} \cite{Le1924} potential is obtained upon setting $\delta_1=12$ and $\delta_2=6$.) For this function:
\[
\deru{\phi}(r)=-\dfrac{c_1\delta_1}{r^{\delta_1+1}}+\dfrac{c_2\delta_2}{r^{\delta_2+1}},\quad
\derd{\phi}(r)=\dfrac{c_1\delta_1(\delta_1+1)}{r^{\delta_1+2}}-\dfrac{c_2\delta_2(\delta_2+1)}{r^{\delta_2+2}},
\]
so that:
\[
\derd{\phi}(r)+\frac{3}{r}\deru{\phi}(r)=
\dfrac{c_1\delta_1(\delta_1-2)}{r^{\delta_1+2}}-
\dfrac{c_2\delta_2(\delta_2-2)}{r^{\delta_2+2}}.
\]
Since $a_A$ es directly proportional to $\sqrt{A}$ (see \eqref{symcpt}), we
have that for \eqref{LJ-potential}, the stability condition \eqref{stab-cond}
holds if and only if $A<A_0$, where $A_0$ is determined from the condition:
\[
\dfrac{c_1\delta_1(\delta_1-2)}{a_A^{\delta_1+2}}-
\dfrac{c_2\delta_2(\delta_2-2)}{a_A^{\delta_2+2}}=0,
\]
from which it follows that:
\begin{equation}\label{LJ-trivBP}
A_0=\frac{\sqrt{3}}{4}\,\left[\frac{c_1\delta_1(\delta_1-2)}{
c_2\delta_2(\delta_2-2)}\right]^{\frac{2}{\delta_1-\delta_2}}.
\end{equation}
Thus \eqref{symcpt} is a (local) solution of \eqref{prob3} if and only
if $A<A_0$. We will show that for $A>A_0$ there exist solutions that break the symmetry $a=b=c$.\qed
\end{ejem}

\begin{figure}
\begin{center}
\scalebox{0.5}{\includegraphics{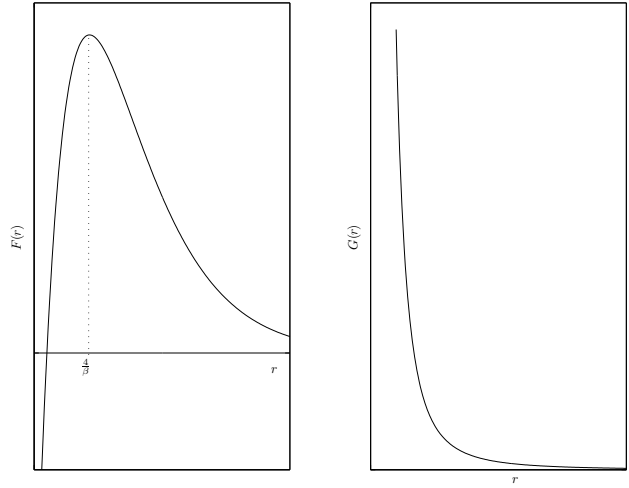}}
\end{center}
\caption{Generic graphs of the functions $F$ and $G$ appearing in the stability
condition \eqref{STBUCK} for a Buckingham potential.}\label{fig:1}
\end{figure}

\begin{ejem}\label{buckA}
A \textit{Buckingham} potential has the form (\cite{Bu1938}, \cite{Te2009}):
\begin{equation}\label{Buck-potential}
\phi(r)=\alpha\me^{-\beta r}-\dfrac{\gamma}{r^\eta},
\end{equation}
where $\alpha,\beta,\gamma,\eta$ are positive constants. Thus
\[
\deru{\phi}(r)=-\alpha\beta\me^{-\beta r}+\dfrac{\gamma\eta}{r^{\eta+1}},\quad
\derd{\phi}(r)=\alpha\beta^2\me^{-\beta
r}-\dfrac{\gamma\eta(\eta+1)}{r^{\eta+2}},
\]
from which it follows that:
\[
\derd{\phi}(r)+\frac{3}{r}\deru{\phi}(r)=
\alpha\beta\left[\beta-\dfrac{3}{r}\right]\me^{-\beta r}-
\dfrac{\gamma\eta(\eta-2)}{r^{\eta+2}}.
\]
After rearrangement, the stability condition \eqref{stab-cond} is equivalent to:
\begin{equation}\label{STBUCK}
F(a_A)>G(a_A),
\end{equation}
where
\[
F(r)=\alpha\beta(\beta r-3)\me^{-\beta r},\quad
G(r)=\dfrac{\gamma\eta(\eta-2)}{r^{\eta+1}}.
\]
These functions generically look as in Figure \ref{fig:1} where for $G$ we
assumed that $\eta>2$. Now clearly $F(r)<G(r)$ for $r$ sufficiently large. Thus
generically we expect the set of values of $A$ for which \eqref{STBUCK} is
satisfied to be of the form $(A_0,A_1)$. Since $F$ has a maximum at
$r_m=\frac{4}{\beta}$, a sufficient condition for this is that $F(r_m)>G(r_m)$,
or after rearrangement, that the coefficients and exponents in
\eqref{Buck-potential} satisfy:
\begin{equation}\label{suffstabcond}
\alpha\beta\me^{-4}>\gamma\eta(\eta-2)\left(\dfrac{\beta}{4}\right)^{\eta+1}.
\end{equation}
To check this condition against the results in \cite{Te2009}, we let
$D,R,\xi>0$ with $\xi>\eta$, and define
\begin{equation}\label{normBUCK}
\alpha=\dfrac{D\eta\me^{\xi}}{\xi-\eta},\quad \beta=\dfrac{\xi}{R},\quad
\gamma=\dfrac{D\xi
R^\eta}{\xi-\eta}.
\end{equation}
It follows that \eqref{Buck-potential} is now given in terms of
$D,R,\xi$ by:
\[
\phi(r)=D\left[\dfrac{\eta}{\xi-\eta}\,\me^{\xi\left(1-\frac{r}{R}\right)}-
\dfrac{\xi}{\xi-\eta}\,\left(\dfrac{R}{r}\right)^{\eta}\right].
\]
Is easy to check now that provided $\xi>\eta+1$, then $\phi$ has negative
minimum value at $r=R$. The results in \cite[Table 1, Page 202]{Te2009} show
that the best fit of a normalized Buckingham potential to a normalized
Lennard--Jones(12-6) potential ($\delta_1=12$ and $\delta_2=6$ in
\eqref{LJ-potential}) is achieved for $\xi=14.3863$ and $\eta=5.6518$. For these
values one can check that \eqref{normBUCK} satisfy the inequality
\eqref{suffstabcond} independent of the values of $D$ and $R$.
\end{ejem}

\begin{ejem}
Consider a potential of the form
\begin{equation}\label{non_spring}
\phi(r)=\frac{1}{2}\,kr^2+\frac{1}{4}\,\beta r^4.
\end{equation}
This potential corresponds to a Hook--type spring when $\beta=0$, a
\textit{hard} spring if $\beta>0$, and a \textit{soft} spring if $\beta<0$. For
this potential
\[
\derd{\phi}(r)+\frac{3}{r}\,\deru{\phi}(r)=4k+6\beta r^2.
\]
Note that the stability condition \eqref{stab-cond} holds when $\beta\ge0$
independent of the value of $A$! That is, the symmetric state \eqref{symcpt} is
a minimizer for all values of $A$. In the case $\beta=0$ is easy to show that
this state is a global minimizer.

On the other hand, if $\beta<0$, the stability condition holds if and only if
$A<A_0$
where
\[
A_0=-\dfrac{k}{2\sqrt{3}\,\beta}.
\]\hspace{\fill}\qed
\end{ejem}

\subsection{Existence and stability of nontrivial solutions}\label{sec:nont3p}
We say that solutions of \eqref{prob3molKT} are \textit{trivial} if
$a=b=c$ and call the set
\begin{equation}\label{triv_branch}
\mc{T}=\set{(\lambda_A,a_A,a_A,a_A,A)\,:\,\lambda_A,a_A\mbox{ given by
\eqref{symcpt}},\, A>0},
\end{equation}
the \textit{trivial branch} parametrized by $A$. In this section we show that
there exist nontrivial solutions of \eqref{prob3molKT} that bifurcate from the
trivial branch.

If we let $\ts{x}=(\lambda,a,b,c)$ and
$\ts{G}\,:\,\Real\times(0,\infty)^4\To\Real^4$ be the left hand side of
\eqref{prob3:1stord}, then this system is equivalent to
$\ts{G}(\ts{x},A)=\ts{0}$. An easy calculation gives that
\begin{equation}\label{ext-hessian}
\Dif_{\ts{x}}\ts{G}(\ts{x},A)=\left[\begin{array}{cc}0&(\Nabla g)^t\\\Nabla g&
\nabla^2E+\lambda\nabla^2g\end{array}\right].
\end{equation}
If we evaluate now at the trivial branch \eqref{symcpt}, we get that
\[
\Dif_{\ts{x}}\ts{G}(\ts{v}_A,A)=\left[\begin{array}{cccc}
0&\gamma&\gamma&\gamma\\\gamma&\alpha&\beta&\beta\\
\gamma&\beta&\alpha&\beta\\\gamma&\beta&\beta&\alpha\end{array}
\right],
\]
where $\ts{v}_A=(\lambda_A,a_A,a_A,a_A)$ and
\[
\alpha=\derd{\phi}(a_A)-\frac{\lambda_A a_A^2}{4},\quad
\beta=\frac{\lambda_A a_A^2}{2},\quad
\gamma=\frac{a_A^3}{4}.
\]
The eigenvalues of this matrix are $\alpha-\beta$, which is a double
eigenvalue with geometric multiplicity two, and the simple eigenvalues
\[
\frac{1}{2}\left[\alpha+2\beta\pm\sqrt{(\alpha+2\beta)^2+12\gamma^2}\right],
\]
which are always nonzero. A pair of linearly independent eigenvectors
corresponding to $\alpha-\beta$ is $\set{(0,-1,1,0)^t,(0,-1,0,1)^t}$.  Since
\[
\alpha-\beta=\derd{\phi}(a_A)-\frac{3}{4}\lambda_A a_A^2=
\derd{\phi}(a_A)+\frac{3}{a_A}\deru{\phi}(a_A),
\]
the double eigenvalue $\alpha-\beta$ becomes zero exactly at the value $A_0$
where the stability condition \eqref{stab-cond} fails by becoming zero. Thus
according to standard theory of bifurcation theory, we can have either none, two
or four branches of solutions of \eqref{prob3:1stord} bifurcating at the point
where $A=A_0$. We now show that there are exactly four branches bifurcating from
such a point: the trivial branch and three branches corresponding to isosceles
triangles.

To avoid the complications of dealing with the two dimensional
kernel of \eqref{ext-hessian} when evaluated at the
trivial branch at $A=A_0$, we make use of the symmetries posses by the mapping
$\ts{G}$. In particular, if we denote by $\mc{G}$ the subgroup of
$\Real^{4\times4}$ of permutations that permute just the $a.b.c$ components of
any $\ts{x}=(\lambda,a,b,c)\in\Real^4$, then
\begin{equation}\label{syms3P}
\ts{G}(P\ts{x},A)=P\ts{G}(\ts{x},A),\quad\forall\, P\in\mc{G}.
\end{equation}
Note that every permutation in $\mc{G}$ changes the eigenvectors
$\set{(0,-1,1,0)^t,(0,-1,0,1)^t}$ of $\alpha-\beta$. However, the eigenvector
\[
\ts{v}\equiv(0,-2,1,1)^t=(0,-1,1,0)^t+(0,-1,0,1)^t,
\]
is unchanged by the proper subgroup of permutations $\mc{H}$ of $\mc{G}$ that
permutes just the $b,c$ components of any $\ts{x}=(\lambda,a,b,c)\in\Real^4$.
Thus $\mc{H}$ is the isotropy subgroup of $\mc{G}$ at $\ts{v}$. The $\mc{H}$
fixed point set is given by:
\[
\Real^4_{\mc{H}}=\set{(\lambda,a,b,b)^t\,:\,\lambda,a,b\in\Real}.
\]
The projection $\mathbb{P}_\mc{H}:\Real^4\To\Real^4_{\mc{H}}$ has matrix
representation:
\[
\mathbb{P}_\mc{H}=\left[\begin{array}{cccc}1&0&0&0\\0&1&0&0\\0&0&\frac{1}{2}
&\frac{1}{2}\\
0&0&\frac{1}{2}&\frac{1}{2}\end{array}\right],
\]
and the $\mc{H}$ reduced problem is now:
\[
\ts{G}_\mc{H}(\ts{u},A)\equiv\mathbb{P}_\mc{H}\ts{G}(\ts{u},A)=\ts{0},\quad
(\ts{u},A)\in\Real^4_{\mc{H}}\times(0,\infty).
\]
Since $\mc{T}\subset\Real^4_{\mc{H}}\times(0,\infty)$, it follows that $\mc{T}$
is a branch of solutions for the $\mc{H}$ reduced problem. Also, since
\[
\Dif_{\ts{u}}\ts{G}_\mc{H}(\ts{u},A)=\mathbb{P}_\mc{H}\Dif_{\ts{x}}\ts{G}(\ts{u}
, A)\mathbb{P}_\mc{H}.
\]
we have that $L_{\mc{H}}(A)\,:\,\Real^4_{\mc{H}}\To\Real^4_{\mc{H}}$ is given
by:
\[
L_{\mc{H}}(A)=\Dif_{\ts{u}}\ts{G}_\mc{H}(\ts{v}_A,A)=
\mathbb{P}_\mc{H}\Dif_{\ts{x}}\ts{G}(\ts{v}_A,A)\mathbb{P}_\mc{H}.
\]
Let
\begin{equation}\label{eigeqn3P}
\mu(A)=\alpha-\beta=\derd{\phi}(a_A)+\frac{3}{a_A}\deru{\phi}(a_A).
\end{equation}
We now have the result for the existence of bifurcating branches for the
reduced problem.
\begin{thm}\label{ntriv3}
Let $\mu(A_0)=0$ and assume that
\begin{equation}\label{TC3p}
\td{\mu}{A}(A_0)\ne0.
\end{equation}
Consider the system \eqref{prob3molKT} and its trivial branch of solutions
\eqref{triv_branch}. Then from the point $(\lambda_0,a_0,a_0,a_0)\in\mc{T}$
bifurcate three branches of nontrivial solutions of \eqref{prob3molKT} each
corresponding to isosceles triangles.
\end{thm}
\begin{prueba}
With the definitions and notation as above, a lengthy but otherwise elementary
calculation shows that for any $A>0$, $\mu(A)$ is a simple eigenvalue of
$L_{\mc{H}}(A)$ restricted to $\Real^4_{\mc{H}}$ with corresponding 
eigenvector $\ts{v}=(0,-2,1,1)^t$. Thus
\[
L_{\mc{H}}(A)\ts{v}=\mu(A)\ts{v},\quad\forall\, A>0.
\]
In particular $\ker(L_{\mc{H}}(A_0))=\Span\{\ts{v}\}$. If we
differentiate with respect to $A$ in the equation above and set $A=A_0$, we get
that
\[
L^\prime_{\mc{H}}(A_0)\ts{v}=\deru{\mu}(A_0)\ts{v}.
\]
Since $L_{\mc{H}}(A_0)$ is symmetric, we have that
$\range(L_{\mc{H}}(A_0))=
\set{\ts{y}\in\Real^n_{\mc{H}}\,:\,\langle\ts{v},\ts{y}\rangle=0}$. Thus the
hypotheses in Theorem \ref{Kras} are satisfied if and only if
$\deru{\mu}(A_0)\ne0$. Thus we get a branch of solutions of the reduced
problem, equivalently \eqref{prob3:1stord}, bifurcating from the trivial branch
at the point where $A=A_0$. Since this branch belongs to
$\Real^4_{\mc{H}}\times\Real$, we can use \eqref{syms3P} to get that there
exist two additional branches of solutions, one belonging to
$\set{(\lambda,b,a,b,A)^t:\lambda,a,b,A\in\Real}$
and the other in $\set{(\lambda,b,b,a,A)^t:\lambda,a,b,A\in\Real}$.
\end{prueba}

\section{Four particles in a tetrahedron}\label{tet:sec}
We now consider the case of four particles arranged in a tetrahedron $T$. Let
$a$, $b$, $c$, $A$, $B$, $C$ be the distances between the
particles where $a,b,c$ denote the lengths of the edges joining a vertex
of $T$, $A$ the length of the edge opposite to $a$, $B$ the length of the
edge opposite to $b$, and $C$ the length of the edge opposite to $c$. The
six--tuple $\ts{a}=(a,b,c,A,B,C)^t$ generates a tetrahedron (\cite{WiDr2009}) if
and only if
\begin{equation}\label{NC_tetrah}
g(\ts{a})>0,\quad A<B+C,\quad B<A+C,\quad C<A+B,
\end{equation}
where $g(\ts{a})$ is given by the Cayley--Menger determinant:
\begin{equation}\label{cm_det}
g(\ts{a})=\left|\begin{array}{ccccc}
0&a^2&b^2&c^2&1\\
a^2&0&C^2&B^2&1\\
b^2&C^2&0&A^2&1\\
c^2&B^2&A^2&0&1\\
1&1&1&1&0
\end{array}\right|.
\end{equation}
If we let $\Real_+$ denote the set of positive real numbers, then we define
\[
\mc{S}=\set{\ts{a}=(a,b,c,A,B,C)^t\in\Real_+^6\,:\,\mbox{\eqref{NC_tetrah}
holds}}.
\]
Note that $\mc{S}$ is open in $\Real^6$. Moreover, any \textit{regular}
tetrahedron in which $a=b=c=A=B=C>0$, is contained in $\mc{S}$.

If $\ts{a}=(a,b,c,A,B,C)^t$ generates a tetrahedron, then so does $P\ts{a}$
where $P=RQ$ and
\begin{enumerate}[i)]
\item
$R$ permutes $(a,b,c)$ and $(A,B,C)$ with the same permutation of three
elements;
\item
$Q$ is any permutation of $(a,b,c,A,B,C)$ in which the base of the
tetrahedron is changed to another face. For example $(c,A,B,C,a,b)$
corresponds to reorienting the tetrahedron so that the base is given by
$(C,a,b)$.
\end{enumerate}
Since there are six permutations of the type $R$ and four of the type $Q$, we
get that there are $24$ permutations of the form $P=RQ$. These $24$
permutations form a subgroup $\mc{R}$ of the group of permutations of six
letters. Also is easy to show that
\begin{equation}\label{CM_sym}
g(P\ts{a})=g(\ts{a}),\quad \forall\,P\in\mc{R}.
\end{equation}
As the Cayley--Menger determinant is directly proportional to the square of the
volume of the tetrahedron (cf. \eqref{prob4}), this identity simply states that
the volume of the tetrahedron remains the same after rotations of the base and
independent of which face we use as the base.

The total energy of the the system of four particles is given now by:
\begin{equation}\label{energy4}
E(\ts{a})=\phi(a)+\phi(b)+\phi(c)+\phi(A)+\phi(B)+\phi(C),
\end{equation}
where the inter--molecular potential $\phi$ is as before. For any $V>0$ we
consider the constrained minimization problem:
 \begin{equation}\label{prob4}
\left\{\begin{array}{c}\displaystyle\min_{\mc{S}}\,E(\ts{a})\\\mbox{subject to
}g(\ts{a})=288V^2.
\end{array}\right.
\end{equation}
The constraint here specifies that the tetrahedron determined by
$\ts{a}$ has volume $V$. The first order necessary conditions for a solution of
this problem are given by\footnote{Since the inequality constraints in the
definition of the set $\mc{S}$ are strict (non--active), the multipliers
corresponding to these constraints are zero.}:
\begin{equation}\label{prob4molKT}
\left\{\begin{array}{rcl} g(\ts{a})-288V^2&=&0,\\\Nabla E(\ts{a})+\lambda\Nabla
g(\ts{a})&=&\ts{0},
\end{array}\right.
\end{equation}
which is now a nonlinear system for the seven unknowns $(\lambda,\ts{a})$ in
terms of the parameter $V$.
\subsection{Existence and stability of trivial states}\label{sec:TS4p}
Expanding the determinant in \eqref{cm_det} and computing its partial
derivatives, we get that
\begin{eqnarray*}
\Nabla g(\ts{a})&=&4\big[
a(A^2(b^2+c^2+B^2+C^2-2a^2-A^2)+(b^2-c^2)(B^2-C^2)),\\
&&~~b(B^2(a^2+c^2+A^2+C^2-2b^2-B^2)+(a^2-c^2)(A^2-C^2)),
\\
&&~~c(C^2(a^2+b^2+A^2+B^2-2c^2-C^2)+(a^2-b^2)(A^2-B^2)),
\\
&&~~A(a^2(b^2+c^2+B^2+C^2-2A^2-a^2)-(b^2-C^2)(c^2-B^2)),
\\
&&~~B(b^2(a^2+c^2+A^2+C^2-2B^2-b^2)-(a^2-C^2)(c^2-A^2)),
\\
&&~~C(c^2(a^2+b^2+A^2+B^2-2C^2-c^2)-(a^2-B^2)(b^2-A^2))
\big]
\end{eqnarray*}
Since $\Nabla
E(\ts{a})=(\deru{\phi}(a),\deru{\phi}(b),\deru{\phi}(c),\deru{\phi}(A),
\deru{\phi}(B),\deru{\phi}(C))^t$, the system \eqref{prob4molKT} when
evaluated at the regular tetrahedron $\ts{a}=(a,a,a,a,a,a)$, reduces to
\[
4a^6=288V^2,\quad \deru{\phi}(a)+4\lambda a^5=0.
\]
This proves the following result.
\begin{lem}
For any $V>0$ the system \eqref{prob4molKT} has the solution
$(\lambda_V,\ts{a}_V,V)$ where $\ts{a}_V=(a_V,a_V,a_V,a_V,a_V,a_V)$ and
\begin{equation}\label{triv_sol_4p}
a_V^3=6\sqrt{2}\,V,\quad\lambda_V=-\dfrac{\deru{\phi}(a_V)}{4a_V^5}.
\end{equation}
\end{lem}
We now examine the stability of the trivial state \eqref{triv_sol_4p}. A lengthy
but otherwise elementary calculation shows that
\begin{equation}\label{hess_triv_4p}
H_V\equiv[\nabla^2E+\lambda_V\nabla^2g](\ts{a}_V)=\left[\begin{array}{cccccc}
\alpha&\beta&\beta&0&\beta&\beta\\
\beta&\alpha&\beta&\beta&0&\beta\\
\beta&\beta&\alpha&\beta&\beta&0\\
0&\beta&\beta&\alpha&\beta&\beta\\
\beta&0&\beta&\beta&\alpha&\beta\\
\beta&\beta&0&\beta&\beta&\alpha
\end{array}\right],
\end{equation}
where
\[
\alpha=\derd{\phi}(a_V)+\dfrac{3}{a_V}\deru{\phi}(a_V),\quad
\beta=-\dfrac{2}{a_V}\deru{\phi}(a_V).
\]
Since $\Nabla g(\ts{a}_V)=4a_V^5(1,1,1,1,1,1)^t$, we need examine the structure
of $H_V$ on the subspace of $\Real^6$ given by
\[
\mc{M}=\set{\ts{y}\in\Real^6\,:\,y_1+y_2+y_3+y_4+y_5+y_6=0}.
\]
We have now the following result:
\begin{thm}\label{triv_state_stab}
Let $\phi\,:\,(0,\infty)\To\Real$ be twice continuously differentiable
function. Then the matrix \eqref{hess_triv_4p} is positive definite over
$\mc{M}$ if and only if $\alpha>0$ and $\alpha-2\beta>0$, which in turn are
equivalent to
\begin{equation}\label{stab-cond_4p}
\derd{\phi}(a_V)+\frac{3}{a_V}\deru{\phi}(a_V)>0,\quad
\derd{\phi}(a_V)+\frac{7}{a_V}\deru{\phi}(a_V)>0.
\end{equation}
Thus the regular tetrahedron $\ts{a}_V$ is a relative minimizer
for the problem \eqref{prob4} for those values of $V$ for
which conditions \eqref{stab-cond_4p} hold.
\end{thm}
\begin{prueba}
Is easy to check that $\mc{M}=\mbox{range}(M)$ where
\[
M=\left[\begin{array}{rrrrr}
1&0&0&0&0\\0&1&0&0&0\\0&0&1&0&0\\-1&-1&-1&-1&-1\\0&0&0&1&0\\
0&0&0&0&1\end{array}\right].
\]
The matrix corresponding to the quadratic form of $H_V$ restricted to $\mc{M}$
is now given by $U_V=M^tH_VM$. The eigenvalues of $U_V$ are:
\[
\alpha\mbox{ (double)},\quad\alpha-2\beta,\quad
\dfrac{1}{2}\left[7\alpha-6\beta\pm\sqrt{16\alpha^2+9(\alpha-2\beta)^2}\right].
\]
Since the product of the last two of these eigenvalues is
$6\alpha(\alpha-2\beta)$, and $7\alpha-6\beta=3(\alpha-2\beta)+4\alpha$, we can
conclude now that they are all positive if
and only if $\alpha>0$ and $\alpha-2\beta>0$. Thus $H_V$ restricted to $\mc{M}$
is positive definite provided these two conditions hold, which in turn implies
that $\ts{a}_V$ is a relative minimizer for problem \eqref{prob4}. That
$\alpha>0$ and $\alpha-2\beta>0$ are equivalent to \eqref{stab-cond_4p} follows
from the definitions of $\alpha$ and $\beta$.
\end{prueba}

\begin{ejem}
For the Lennard--Jones potential \eqref{LJ-potential}
we have that:
\begin{eqnarray*}
\derd{\phi}(r)+\frac{3}{r}\deru{\phi}(r)&=&
\dfrac{c_1\delta_1(\delta_1-2)}{r^{\delta_1+2}}-
\dfrac{c_2\delta_2(\delta_2-2)}{r^{\delta_2+2}},\\
\derd{\phi}(r)+\frac{7}{r}\deru{\phi}(r)&=&
\dfrac{c_1\delta_1(\delta_1-6)}{r^{\delta_1+2}}-
\dfrac{c_2\delta_2(\delta_2-6)}{r^{\delta_2+2}}
\end{eqnarray*}
For simplicity, we assume $\delta_1>6$. We now have two cases:
\begin{enumerate}[i)]
\item
Assume that $\delta_2\in(2,6]$. Then the second condition in
\eqref{stab-cond_4p} is automatically satisfied
and the first condition holds if and only if $V<V_0$, where $V_0$ is determined
from the condition (cf. \eqref{triv_sol_4p}):
\[
\dfrac{c_1\delta_1(\delta_1-2)}{r_0^{\delta_1+2}}-
\dfrac{c_2\delta_2(\delta_2-2)}{r_0^{\delta_2+2}}=0,\quad r_0=a_{V_0},
\]
from which it follows that:
\[
V_0=\dfrac{\sqrt{2}}{12}\left[\frac{c_1\delta_1(\delta_1-2)}{
c_2\delta_2(\delta_2-2)}\right]^{\frac{3}{\delta_1-\delta_2}}.
\]
Thus in this case the regular tetrahedron $\ts{a}_V$ is a (local) solution of
\eqref{prob4} if and only if $V<V_0$.
\item
If $\delta_2>6$, then the second condition in \eqref{stab-cond_4p} holds if and
only if $V<V_1$, where $V_1$ is determined
from the condition:
\[
\dfrac{c_1\delta_1(\delta_1-6)}{r_1^{\delta_1+2}}-
\dfrac{c_2\delta_2(\delta_2-6)}{r_1^{\delta_2+2}}=0,\quad r_1=a_{V_1},
\]
from which it follows that:
\[
V_1=\dfrac{\sqrt{2}}{12}\left[\frac{c_1\delta_1(\delta_1-6)}{
c_2\delta_2(\delta_2-6)}\right]^{\frac{3}{\delta_1-\delta_2}}.
\]
Since $\delta_1>\delta_2>6$, it follows that $V_1<V_0$. Thus in this case the
regular tetrahedron $\ts{a}_V$ is a (local) solution of
\eqref{prob4} if and only if $V<V_1$.\hspace{\fill}\qed
\end{enumerate}
\end{ejem}

\begin{ejem}
For the Buckingham \eqref{Buck-potential}, we have that
\begin{eqnarray*}
\derd{\phi}(r)+\frac{3}{r}\deru{\phi}(r)&=&
\alpha\beta\left[\beta-\dfrac{3}{r}\right]\me^{-\beta r}-
\dfrac{\gamma\eta(\eta-2)}{r^{\eta+2}},\\
\derd{\phi}(r)+\frac{7}{r}\deru{\phi}(r)&=&
\alpha\beta\left[\beta-\dfrac{7}{r}\right]\me^{-\beta r}-
\dfrac{\gamma\eta(\eta-6)}{r^{\eta+2}}.
\end{eqnarray*}
Note that the first condition in \eqref{stab-cond_4p} holds for an interval
$(V_0,V_1)$ of volume values under the conditions \eqref{suffstabcond} in
Example \ref{buckA}. The analysis now becomes rather complicated and we
just describe it qualitatively. If $\eta>6$, then the second condition in
\eqref{stab-cond_4p} would hold as well for values of $V$ in an interval of the
form $(V_2,V_3)$ provided some condition similar to \eqref{suffstabcond} holds.
Depending as to whether or not the intersection $(V_0,V_1)\cap(V_2,V_3)$
is non--empty, we might get stable regular tetrahedrons. On the other hand, if
$\eta\in(2,6]$, then  the second condition in \eqref{stab-cond_4p} would hold as
well for values of $V$ in an interval of the form $(V_4,\infty)$ and again the
existence of trivial states will depend on whether the corresponding
intersection is non--empty.
\end{ejem}

\begin{ejem}
For the potential \eqref{non_spring}
\[
\derd{\phi}(r)+\frac{3}{r}\,\deru{\phi}(r)=4k+6\beta r^2,\quad
\derd{\phi}(r)+\frac{7}{r}\,\deru{\phi}(r)=8k+10\beta r^2.
\]
Note that the stability conditions \eqref{stab-cond_4p} holds when $\beta\ge0$
independent of the value of $V$! That is, the regular tetrahedron $\ts{a}_V$ is
a relative minimizer for the problem \eqref{prob4} for all values of
$V$. In the case $\beta=0$, since the functional \eqref{energy4} is convex,
this state is a global minimizer.

On the other hand, if $\beta<0$, the first condition in \eqref{stab-cond_4p}
holds if $V<V_1$ and the second condition if $V<V_2$ where
\[
V_1^2=-\dfrac{k^3}{243\,\beta^3},\quad V_2^2=-\dfrac{8k^3}{1125\,\beta^3}.
\]
Since $V_1<V_2$ we get that conditions \eqref{stab-cond_4p} hold both for
$V<V_1$. For $V>V_1$ either one or both conditions fail.\hspace{\fill}\qed
\end{ejem}

\subsection{Existence and stability of nontrivial states}\label{sec:NS4p}
Let $\ts{G}\,:\,\Real\times\mc{S}\times(0,\infty)\To\Real^7$ be given by the
left hand side of \eqref{prob4molKT}:
\[
\ts{G}(\ts{x},V)=\left[\begin{array}{c}g(\ts{a})-288V^2\\
\Nabla E(\ts{a})+\lambda\Nabla g(\ts{a})\end{array}\right],
\]
where $\ts{x}=(\lambda,\ts{a})$. We have now that
\begin{equation}\label{ext4-hessian}
\Dif_{\ts{x}}\ts{G}(\ts{x},V)=\left[\begin{array}{cc}0&(\Nabla
g(\ts{a}))^t\\\Nabla g(\ts{a})&
\nabla^2E(\ts{a})+\lambda\nabla^2g(\ts{a})\end{array}\right].
\end{equation}
It follows from \eqref{CM_sym} that
\begin{subequations}\label{pb4_sym}
\begin{eqnarray}
\Nabla g(P\ts{a})&=&P\Nabla g(\ts{a}),\label{grad_sym}\\
\nabla^2g(P\ts{a})&=&P\nabla^2g(\ts{a})P^t,\quad P\in\mc{R}, \label{hess_sym}
\end{eqnarray}
\end{subequations}
with similar relations for the total energy $E$. It follows now from
\eqref{grad_sym} that
\begin{equation}\label{KT4_sym}
\ts{G}(Q\ts{x},V)=Q\ts{G}(\ts{x},V),\quad Q\in\mc{G},
\end{equation}
where
\[
\mc{G}=\set{Q=\left[\begin{array}{cc}1&\ts{0}^t\\\ts{0}&P\end{array}\right]\,:\,
\ P\in\mc{R}}.
\]
Thus the system \eqref{prob4molKT} remains the same, up to reordering of the
equations, when $\ts{x}=(\lambda,\ts{a})$ is replaced by $Q\ts{x}$.

We now begin the analysis of the existence of solutions of the system
\eqref{prob4molKT} bifurcating from the trivial branch:
\[
\mc{T}=\set{(\lambda_V,\ts{a}_V,V)\,:\,\lambda_V,\,\ts{a}_V\mbox{ given by
\eqref{triv_sol_4p}},\, V>0}.
\]
If we evaluate \eqref{ext4-hessian} at the trivial state
$(\lambda_V,\ts{a}_V,V)$, then this matrix reduces to:
\begin{equation}\label{ext4-hessian_triv}
\Dif_{\ts{x}}\ts{G}(\lambda_V,\ts{a}_V,V)=\left[\begin{array}{cc}0&(\Nabla
g(\ts{a}_V))^t\\\Nabla g(\ts{a}_V)&
H_V\end{array}\right],
\end{equation}
where $\Nabla g(\ts{a}_V)=4a_V^5(1,1,1,1,1,1)^t$ and $H_V$ is given by
\eqref{hess_triv_4p}. The matrix \eqref{ext4-hessian_triv} has two eigenvalues
which are nonzero for every value of $V>0$, with the remaining eigenvalues
given by:
\begin{enumerate}
\item
$\mu_1(V)=\derd{\phi}(a_V)+\frac{3}{a_V}\deru{\phi}(a_V)$ with algebraic and
geometric multiplicity three, and corresponding eigenvectors:
\begin{equation}\label{ev_mu1}
(0,-1,0,0,1,0,0)^t,\quad(0,0,-1,0,0,1,0)^t,\quad (0,0,0,-1,0,0,1)^t.
\end{equation}
\item
$\mu_2(V)=\derd{\phi}(a_V)+\frac{7}{a_V}\deru{\phi}(a_V)$ with algebraic and
geometric multiplicity two, and corresponding eigenvectors:
\begin{equation}\label{ev_mu2}
(0,-1,1,0,-1,1,0)^t,\quad(0,-1,0,1,-1,0,1)^t.
\end{equation}
\end{enumerate}
\begin{remark}
Note that the expressions for these eigenvalues are the ones that appear in
Theorem \ref{triv_state_stab} characterizing the stability of the trivial state
$(\lambda_V,\ts{a}_V,V)$. Thus the trivial state can change stability exactly
when one of these two eigenvalues becomes zero.
\end{remark}

To deal with these kernels with dimension greater than one, we proceed as in the
previous section by considering a suitable reduced problem in each case. These
reductions are determined by the symmetries present in this problem which are
embodied in \eqref{KT4_sym}.
\subsubsection{The eigenvalue $\mu_1(V)$}
Let us take the eigenvector $\ts{g}=(0,0,0,-1,0,0,1)^t$ of the eigenvalue
$\mu_1(V)$ above. (The analysis for the other two eigenvectors is similar.)
By inspection is easy to get that the isotropy subgroup $\mc{H}$ of $\mc{G}$ at
$\ts{g}$ is given by:
\begin{eqnarray*}
\mc{H}&=&\left\{
\left(\begin{array}{ccccccc}\lambda&a&b&c&A&B&C\\\lambda&a&b&c&A&B&C\end{array}
\right),
\left(\begin{array}{ccccccc}\lambda&a&b&c&A&B&C\\\lambda&b&a&c&B&A&C\end{array}
\right),\right.\\
&&\left.\left(\begin{array}{ccccccc}\lambda&a&b&c&A&B&C\\\lambda&B&A&c&b&a&C
\end{array}\right),
\left(\begin{array}{ccccccc}\lambda&a&b&c&A&B&C\\\lambda&A&B&c&a&b&C\end{array}
\right)\right\}.
\end{eqnarray*}
The $\mc{H}$--\textit{fixed point set} is now given by:
\begin{equation}\label{HFPtmu1a}
\Real^7_{\mc{H}}=\set{(\lambda,a,a,c,a,a,C)^t\,:\,\lambda,a,c,C\in\Real}.
\end{equation}
The projection $\mathbb{P}_\mc{H}:\Real^7\To\Real^7_{\mc{H}}$ has matrix
representation:
\[
\mathbb{P}_\mc{H}=\left[\begin{array}{ccccccc}
 1&   0&   0& 0&   0&   0& 0\\
 0& \frac{1}{4}& \frac{1}{4}& 0& \frac{1}{4}& \frac{1}{4}& 0\\
 0& \frac{1}{4}& \frac{1}{4}& 0& \frac{1}{4}& \frac{1}{4}& 0\\
 0&   0&   0& 1&   0&   0& 0\\
 0& \frac{1}{4}& \frac{1}{4}& 0& \frac{1}{4}& \frac{1}{4}& 0\\
 0& \frac{1}{4}& \frac{1}{4}& 0& \frac{1}{4}& \frac{1}{4}& 0\\
 0&   0&   0& 0&   0&   0& 1\\
\end{array}\right],
\]
and the $\mc{H}$ reduced problem is now:
\[
\ts{G}_\mc{H}(\ts{u},V)\equiv\mathbb{P}_\mc{H}\ts{G}(\ts{u},V)=\ts{0},\quad
(\ts{u},V)\in\Real^7_{\mc{H}}\times(0,\infty).
\]
Since $\mc{T}\subset\Real^7_{\mc{H}}\times(0,\infty)$, it follows that $\mc{T}$
is a branch of solutions for the $\mc{H}$ reduced problem. Also, since
\[
\Dif_{\ts{u}}\ts{G}_\mc{H}(\ts{u},V)=\mathbb{P}_\mc{H}\Dif_{\ts{x}}\ts{G}(\ts{u}
, V)\mathbb{P}_\mc{H}.
\]
we have that $L_{\mc{H}}(V)\,:\,\Real^7_{\mc{H}}\To\Real^7_{\mc{H}}$ is given
by:
\[
L_{\mc{H}}(V)=\Dif_{\ts{u}}\ts{G}_\mc{H}(\lambda_V,\ts{a}_V,V)=
\mathbb{P}_\mc{H}\Dif_{\ts{x}}\ts{G}(\lambda_V,\ts{a}_V,V)\mathbb{P}_\mc{H}.
\]
It easy to check now that $\mu_1(V)$ is a simple eigenvalue of
$L_{\mc{H}}(V)$ restricted to $\Real^7_{\mc{H}}$ with corresponding 
eigenvector $\ts{g}$, that is
\[
L_{\mc{H}}(V)\ts{g}=\mu_1(V)\ts{g},\quad\forall\, V>0.
\]
We now have the result for the existence of bifurcating branches for the
reduced problem. We omit the proof as it is similar to that of Theorem
\ref{ntriv3}.

\begin{thm}\label{ntriv4_b1}
Let $\mu_1(V_1)=0$ and assume that $\td{\mu_1}{V}(V_1)\ne0$. Then the system
\eqref{prob4molKT} has a branch of nontrivial solutions in
$\Real^7_{\mc{H}}\times(0,\infty)$ bifurcating from the trivial branch $\mc{T}$
at the point where $V=V_1$, where $\Real^7_{\mc{H}}$ is given by
\eqref{HFPtmu1a}.
\end{thm}

\begin{remark}
It follows from \eqref{KT4_sym} that there are two additional branches of
nontrivial solutions of the system \eqref{prob4molKT} of the forms:
\begin{eqnarray*}
& \set{(\lambda,a,c,a,a,C,a)\,:\,\lambda\in\Real,\,a,c,C>0},&\\
&\set{(\lambda,c,a,a,C,a,a)\,:\,\lambda\in\Real,\,a,c,C>0}.&
\end{eqnarray*}
\end{remark}

We now consider the case of the eigenvector $\ts{g}=(0,-1,-1,-1,1,1,1)^T$ of
$\mu_1(V)$. This eigenvector is obtained by adding the three eigenvector in
\eqref{ev_mu1}. By inspection, the isotropy subgroup $\mc{H}$ of $\mc{G}$ at
$\ts{g}$ is given by those permutations in $\mc{G}$ that permute the symbols
$(a,b,c)$ and $(A,B,C)$ in $(\lambda,a,b,c,A,B,C)$ with the same permutation.
(Thus $\mc{H}$ has six elements.) The $\mc{H}$--fixed point set is now given by:
\begin{equation}\label{FPtmu1b}
\Real^7_{\mc{H}}=\set{(\lambda,a,a,a,A,A,A)^t\,:\,\lambda,a,A\in\Real}.
\end{equation}
The projection $\mathbb{P}_\mc{H}:\Real^7\To\Real^7_{\mc{H}}$ has matrix
representation:
\[
\mathbb{P}_\mc{H}=\left[\begin{array}{ccccccc}
 1&   0&   0& 0&   0&   0& 0\\
 0& \frac{1}{3}& \frac{1}{3}& \frac{1}{3}& 0&0&0\\
 0& \frac{1}{3}& \frac{1}{3}& \frac{1}{3}& 0&0&0\\
 0& \frac{1}{3}& \frac{1}{3}& \frac{1}{3}& 0&0&0\\
 0& 0& 0& 0& \frac{1}{3}& \frac{1}{3}& \frac{1}{3}\\
 0& 0& 0& 0& \frac{1}{3}& \frac{1}{3}& \frac{1}{3}\\
 0& 0& 0& 0& \frac{1}{3}& \frac{1}{3}& \frac{1}{3}\\
\end{array}\right],
\]
It follows now that for the $\mc{H}$ reduced problem, $\mu_1(V)$ is a simple
eigenvalue with corresponding eigenvector $\ts{g}$. The proof of the following
result is as that of Theorem \ref{ntriv3}.

\begin{thm}\label{ntriv4_b2}
Let $\mu_1(V_1)=0$ and assume that $\td{\mu_1}{V}(V_1)\ne0$. Then the system
\eqref{prob4molKT} has a branch of nontrivial solutions in
$\Real^7_{\mc{H}}\times(0,\infty)$ bifurcating from the trivial branch $\mc{T}$
at the point where $V=V_1$, where $\Real^7_{\mc{H}}$ is given by
\eqref{FPtmu1b}.
\end{thm}

\begin{remark}
By applying all the transformations in $\mc{G}$, it follows from
\eqref{KT4_sym} that there are three additional branches of solutions of the
system \eqref{prob4molKT} of the forms:
\begin{eqnarray*}
& \set{(\lambda,A,A,a,a,a,A)\,:\,\lambda\in\Real,\,a,A>0},&\\
&\set{(\lambda,a,A,A,A,a,a)\,:\,\lambda\in\Real,\,a,A>0},&\\
&\set{(\lambda,A,a,A,a,A,a)\,:\,\lambda\in\Real,\,a,A>0}.&
\end{eqnarray*}
Thus combining both theorems, we get that there are seven branches of
nontrivial solutions bifurcating from the trivial branch
$\set{(\lambda_V,\ts{a}_V,V)\,:\,V>0}$ at the value of $V=V_1$ where
$\mu_1(V_1)=0$ and $\deru{\mu_1}(V_1)\ne0$.
\end{remark}

\subsubsection{The eigenvalue $\mu_2(V)$}
We now consider the case of the eigenvector $\ts{g}=(0,-2,1,1,-2,1,1)^T$ of
$\mu_2(V)$. This eigenvector is obtained by adding the eigenvectors in
\eqref{ev_mu2}. By inspection, the isotropy subgroup $\mc{H}$ of $\mc{G}$ at
$\ts{g}$ is given by 
\begin{eqnarray*}
\mc{H}&=&\left\{
\left(\begin{array}{ccccccc}\lambda&a&b&c&A&B&C\\\lambda&a&b&c&A&B&C\end{array}
\right),
\left(\begin{array}{ccccccc}\lambda&a&b&c&A&B&C\\\lambda&a&c&b&A&C&B\end{array}
\right),\right.\\
&&\left.\left(\begin{array}{ccccccc}\lambda&a&b&c&A&B&C\\\lambda&A&C&b&a&c&B
\end{array}\right),
\left(\begin{array}{ccccccc}\lambda&a&b&c&A&B&C\\\lambda&A&b&C&a&B&c\end{array}
\right)\right.\\
&&\left.\left(\begin{array}{ccccccc}\lambda&a&b&c&A&B&C\\\lambda&A&B&c&a&b&C
\end{array}\right),
\left(\begin{array}{ccccccc}\lambda&a&b&c&A&B&C\\\lambda&A&c&B&a&C&b\end{array}
\right)\right\},
\end{eqnarray*}
with $\mc{H}$--fixed point set given by:
\begin{equation}\label{FPtmu2}
\Real^7_{\mc{H}}=\set{(\lambda,a,b,b,a,b,b)\,:\,\lambda,a,b\in\Real}.
\end{equation}
The projection $\mathbb{P}_\mc{H}:\Real^7\To\Real^7_{\mc{H}}$ has matrix
representation:
\[
\mathbb{P}_\mc{H}=\left[\begin{array}{ccccccc}
 1&   0&   0& 0&   0&   0& 0\\
 0& \frac{1}{2}& 0& 0& \frac{1}{2}&0&0\\
 0& 0& \frac{1}{4}& \frac{1}{4}& 0&\frac{1}{4}&\frac{1}{4}\\
 0& 0& \frac{1}{4}& \frac{1}{4}& 0&\frac{1}{4}&\frac{1}{4}\\
 0& \frac{1}{2}& 0& 0& \frac{1}{2}&0&0\\
 0& 0& \frac{1}{4}& \frac{1}{4}& 0&\frac{1}{4}&\frac{1}{4}\\
 0& 0& \frac{1}{4}& \frac{1}{4}& 0&\frac{1}{4}&\frac{1}{4}\\
\end{array}\right],
\]
It follows now that for the $\mc{H}$ reduced problem, $\mu_2(V)$ is a simple
eigenvalue with corresponding eigenvector $\ts{g}$. The proof of the following
result is as that of Theorem \ref{ntriv3}.

\begin{thm}\label{ntriv4_b3}
Let $\mu_2(V_2)=0$ and assume that $\td{\mu_2}{V}(V_2)\ne0$. Then the system
\eqref{prob4molKT} has a branch of nontrivial solutions in
$\Real^7_{\mc{H}}\times(0,\infty)$ bifurcating from the trivial branch $\mc{T}$
at the point where $V=V_2$, where $\Real^7_{\mc{H}}$ is given by
\eqref{FPtmu2}.
\end{thm}

\begin{remark}
By applying all the transformations in $\mc{G}$, it follows from
\eqref{KT4_sym} that there are two additional branches of solutions of the
system \eqref{prob4molKT} of the forms:
\begin{eqnarray*}
& \set{(\lambda,b,a,b,b,a,b)\,:\,\lambda\in\Real,\,a,b>0},&\\
&\set{(\lambda,b,b,a,b,b,a)\,:\,\lambda\in\Real,\,a,b>0}.&
\end{eqnarray*}
\end{remark}

\section{Numerical examples}\label{sec:NE}
In this section we present some numerical examples illustrating the
results of the previous sections. For simplicity we limit ourselves to the
three particle problem. The examples show that the structure of the bifurcation
diagrams is quite rich and complex. To construct the pictures in this section,
we use the results of Theorem \ref{ntriv3}, in particular the symmetries given
by \eqref{syms3P}, together with various numerical techniques to get full
or detailed descriptions of the corresponding bifurcation diagram.

To compute approximations of the bifurcating solutions predicted by Theorem
\ref{ntriv3}, one employs a predictor--corrector continuation method (cf.
\cite{AllGe1979}, \cite{Ke1986}). The bifurcation points off the trivial branch
can be determined, by Theorem \ref{ntriv3}, from the solutions of the equation
$\mu(A)=0$ (cf. \eqref{symcpt}$_1$, \eqref{eigeqn3P}). Secondary bifurcation
points off nontrivial branches can be detected by monitoring the sign of a
certain determinant. Once a sign change in this determinant is detected, the
bifurcation point can be computed by a bisection or secant type iteration. After
detection and computation of a bifurcation point, then one can use formulas
\eqref{bif2}--\eqref{bif4} to get an approximate point on the solution curve
from which the continuation of the bifurcating branch can proceed. 

Any trivial or nontrivial computed solution $(\ts{x}^*,A^*)$ will be called
\textit{stable}, if the matrix $(\nabla^2E+\lambda\nabla^2g)(\ts{x}^*,A^*)$ (cf.
\eqref{ext-hessian}) is positive definite when restricted to the tangent space
at $\ts{x}^*$ of the constraint of fixed area. Otherwise the point
$(\ts{x}^*,A^*)$ will be called \textit{unstable}. We recall that the tangent
space at $\ts{x}^*$ of the constraint of fixed area is given by 
\[
\mc{M}=\set{(x,y,z)\,:\,\Nabla g(\ts{x}^*)\cdot (x,y,z)=0}.
\]
Note that since this space depends on the point $\ts{x}^*$, then except for the
trivial branch where the solution is known explicitly, the stability of a
solution can only be determined numerically. 

\begin{figure}
\begin{center}
\scalebox{0.4}{\includegraphics{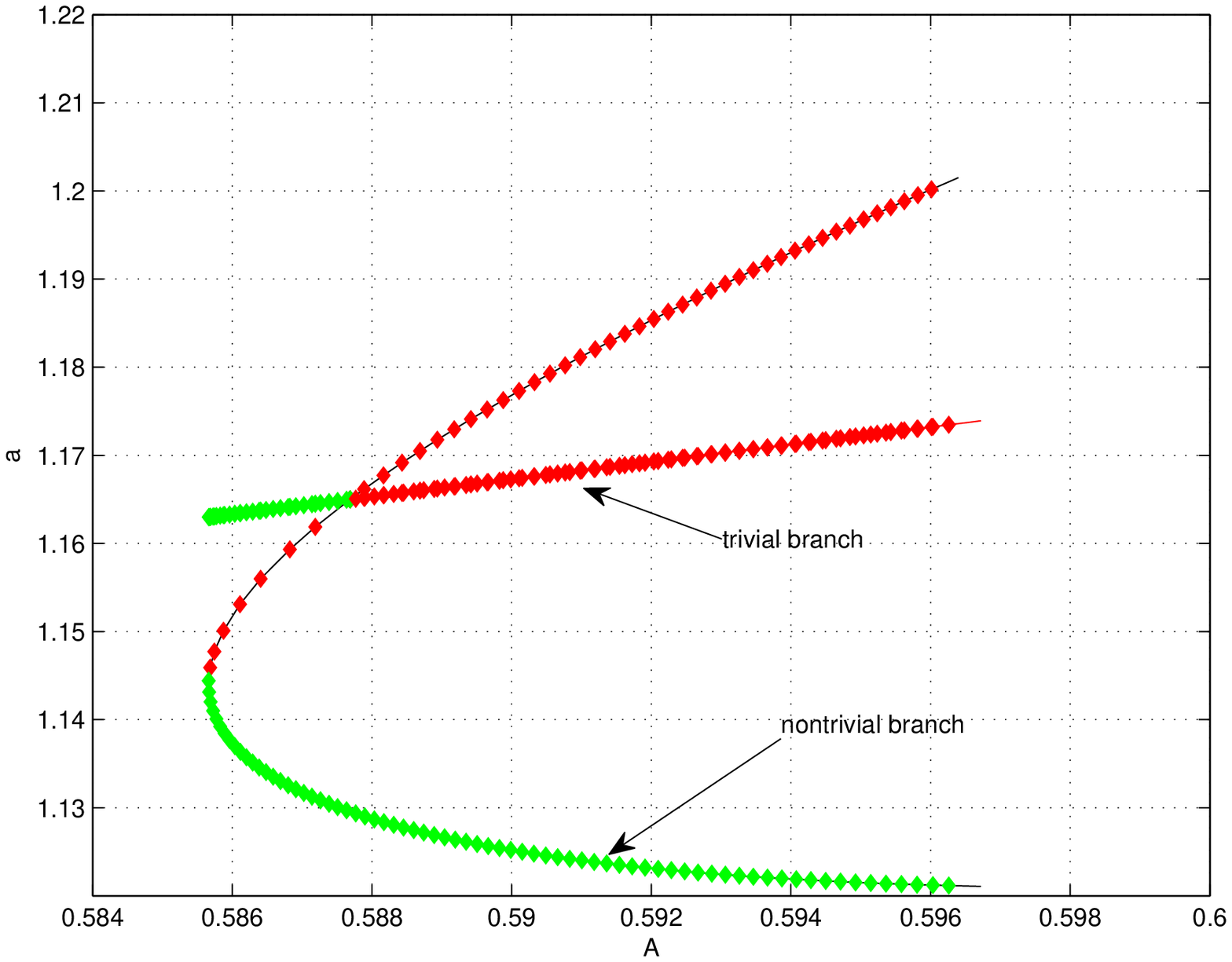}}
\end{center}
\caption{Bifurcation diagram for the $a$ component vs $A$ for the system
\eqref{prob3molKT} in the case $a=b$ and a Lennard-Jones potential. The points
in green represent local minima of \eqref{prob3} while those in red are either
maxima or none.}\label{fig:2}
\end{figure}

\begin{figure}
\begin{center}
\scalebox{0.4}{\includegraphics{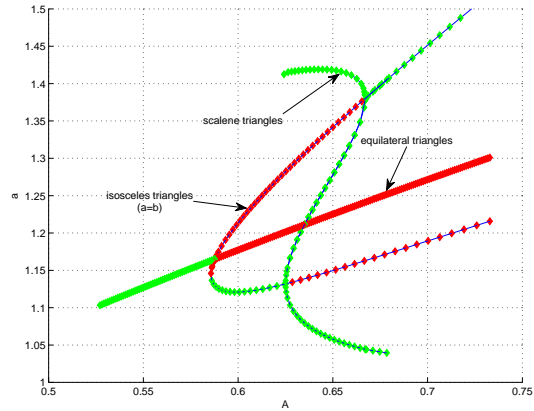}}
\end{center}
\caption{Bifurcation diagram for the system \eqref{prob3molKT} in the case of a
Lennard-Jones potential for a larger interval of values of $A$.
There are secondary bifurcations into stable scalene triangles.}\label{fig:3}
\end{figure}

\begin{figure}
\begin{center}
\scalebox{0.5}{\includegraphics{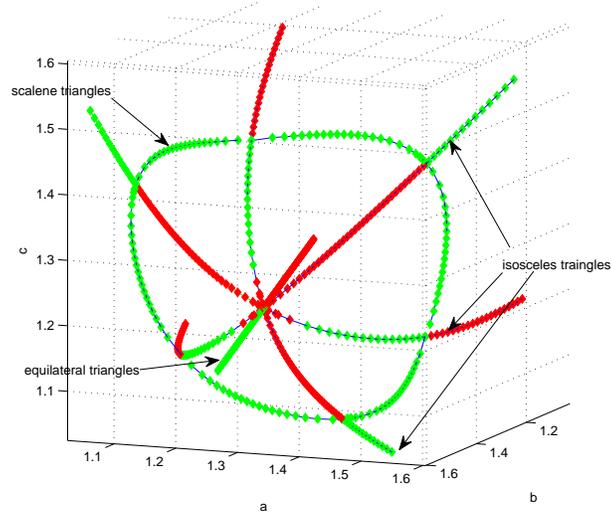}}
\end{center}
\caption{Solution set for the system \eqref{prob3molKT} in the case of a
Lennard-Jones potential without the $A$ dependence.}\label{fig:4}
\end{figure}

\begin{figure}
\begin{center}
\scalebox{0.5}{\includegraphics{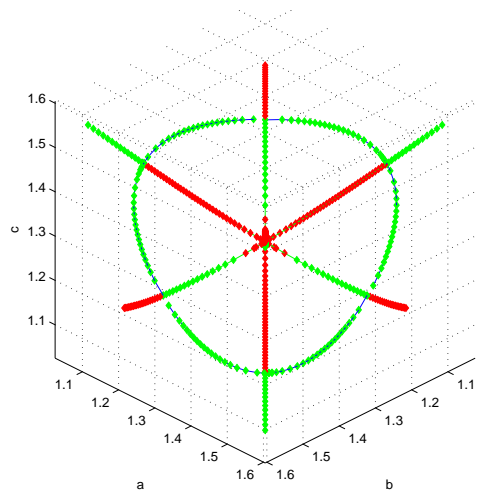}}
\end{center}
\caption{Solution set for the system \eqref{prob3molKT} in the case of a
Lennard-Jones potential without the $A$ dependence with the branch of trivial
solutions coming out of the page.}\label{fig:5}
\end{figure}

In our first example we consider the Lennard-Jones potential 
\eqref{LJ-potential} with $c_1=1$, $c_2=2$, $\delta_1=12$, and
$\delta_2=6$. (We obtained similar results for other values of  $c_1,c_2$
like those for argon in which $c_1/c_2=3.4^6$ \AA{}$^6$.) From equation
\eqref{LJ-trivBP} we get that the bifurcation point off the trivial branch is
given approximately by $A_0=0.5877$. For the case of Theorem \ref{ntriv3} in
which $a=b$, we show in Figure \ref{fig:2} a close up of the bifurcating branch
near this bifurcation point, for the projection onto the $A$--$a$ plane. In this
figure and the others, the color red indicates unstable solutions while the
stable ones are marked in green. Note that the bifurcation is of the
trans--critical type. It is interesting to note that for an interval of values
of the parameter $A$ to the left of $A_0$ in the figure (approximately
$(0.5855,0.5877)$), there are multiple states (trivial and nontrivial) which are
stable, the trivial one having an energy less than the nontrivial state in this
case. In Figure \ref{fig:3} we look at the same branches of solutions, again the
projection onto the $A$--$a$ plane, but for a larger interval of values of $A$.
We now discover that there are two secondary bifurcation points\footnote{In
Figure \ref{fig:3} there are bifurcations only corresponding to the values of
$A_0=0.5877$, $A_1=0.6251$ and $A_2=0.6670$. The apparent crossing of a branch
of scalene triangles and the trivial branch is just an artifact of the
projection onto the $A$--$a$ plane.} at approximately $A_1=0.6251$ and
$A_2=0.6670$, and once again we have multiple stable states (with different
symmetries) existing for an interval of values of the parameter $A$. The
branches of solutions bifurcating at these values of $A$ correspond to stable
scalene triangles. Once a branch of solutions is computed, we can use the
symmetries \eqref{syms3P} to generate other branches of solutions. In Figure
\ref{fig:4} we show all the solutions obtained via this process, projected to
the $abc$ space (no $A$ dependence). Figure \ref{fig:5} show the same set of
solution but with the branch or axis of trivial solutions coming out of the
page. The figures clearly show the variety of solutions (stable an unstable) for
the problem \eqref{prob3} as well as the rather complexity of the corresponding
solution set.

\begin{figure}
\begin{center}
\scalebox{0.6}{\includegraphics{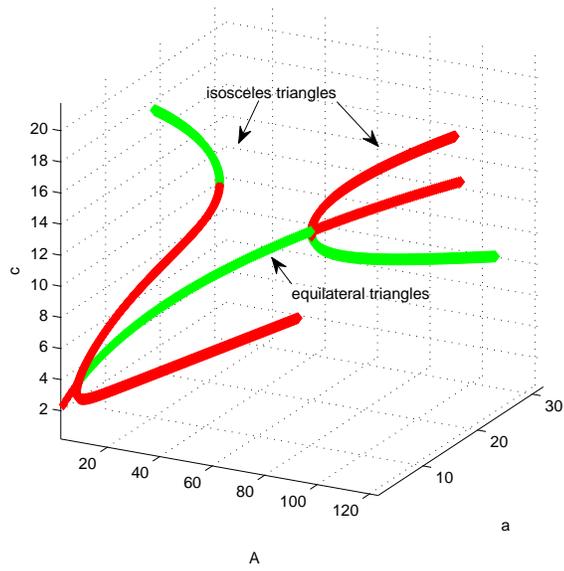}}
\end{center}
\caption{Dependence of the $a$ and $c$ components on the parameter $A$ for the
system \eqref{prob3molKT} in the case $a=b$ and for a Buckingham
potential.}\label{fig:6}
\end{figure}

\begin{figure}
\begin{center}
\scalebox{0.5}{\includegraphics{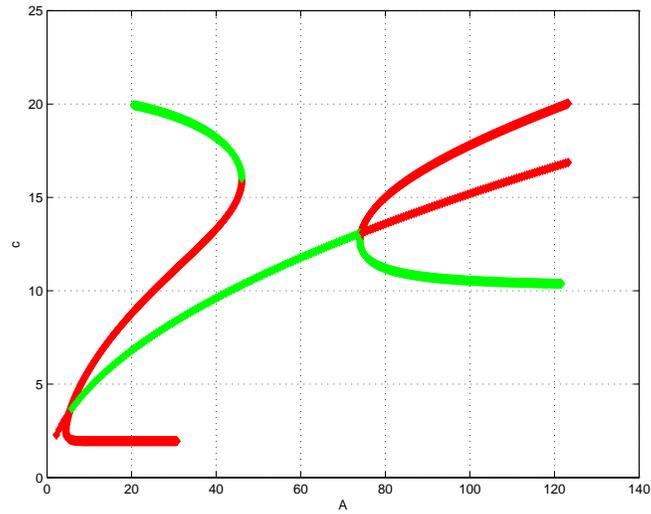}}
\end{center}
\caption{Projection onto the $c$ vs $A$ plane of the set in
Figure \ref{fig:6}.}\label{fig:7}
\end{figure}

For our next numerical example we consider the Buckingham potential
\eqref{Buck-potential} with parameter values $\alpha=\beta=\gamma=1$ and
$\eta=4$, which satisfy \eqref{suffstabcond}. In this case we have two
bifurcation points off the trivial branch (which correspond to solutions of
$\mu(A)=0$) at approximately $A_0=5.3154$ and $A_1=74.2253$. The trivial branch
is stable for $A\in(A_0,A_1)$ and unstable otherwise. Both bifurcations are into
isosceles triangles, and both are of trans--critical type but with different
stability patterns. In Figure \ref{fig:6} we show the solution set for the case
$a=b$. The plot shows the dependence of the $a$ and $c$ components on the area
parameter $A$. In Figure \ref{fig:7} we show the projection of this set onto
the $c$ vs $A$ plane where one can appreciate somewhat better the stability
patterns at each bifurcation point, and that there is a turning point for
$A\approx46$ on the branch bifurcating from $A_0$. Note that once again we have
multiple stable states existing for an interval of values of the parameter $A$.
No secondary bifurcations were detected in this case.

\section{Final comments}
The variety or type of solutions obtained from Theorems \ref{ntriv3},
\ref{ntriv4_b1}, \ref{ntriv4_b2}, and \ref{ntriv4_b3}, could be predicted
generically from an analysis of the symmetries present in our problem as given
by \eqref{syms3P} and \eqref{KT4_sym}. However such an analysis does not
guaranties the existence of solutions with the predicted symmetries. It is the
application of the Equivariant Bifurcation Theorem \ref{Kras} that actually
yields the result that such solutions exist. The generic analysis however
is a preliminary step in identifying the spaces in which Theorem
\ref{Kras} can be applied. We should also point out that the results on the
bifurcating branches in Theorems \ref{ntriv3}, \ref{ntriv4_b1}, \ref{ntriv4_b2},
and \ref{ntriv4_b3} are global in the sense that the so called Crandall and
Rabinowitz alternatives in Theorem \ref{Kras} hold. That is, any bifurcating
branch is either unbounded, or it intersects the boundary of the domain of
definition of the operator in the equilibrium conditions, or it intersects the
trivial branch at another eigenvalue.

The results of this paper might be useful in the analysis of the more general
and complex problem of arrays with many particles. As the total area or volume
of such an array is increased, thus reducing its density, one might expect that
locally situations similar to the ones discussed in this paper might be taking
place in different parts of the array. It is interesting to note here that the
existence of multiple stable configurations detected in the numerical examples
of Section \ref{sec:NE}, opens up the possibility for the existence of multiple
equilibrium (local) states in a large molecular array, reminiscent of the bubble
formation phenomena mentioned in the introduction.\\

\noindent\textbf{Acknowledgements:}
This research was sponsored in part (Negr\'on--Marrero and L\'opez--Serrano) by
the NSF--PREM Program of the UPRH (Grant No. DMR--0934195).

\pagebreak
\end{document}